%% file: main.tex
\documentclass[
aps,
prl,
reprint
]{revtex4-2}

\usepackage{graphicx}
\usepackage{xcolor}
\usepackage{amsmath} 
\usepackage{amssymb}
\usepackage{mathtools}
\usepackage{physics}
\usepackage{times, multirow, amsfonts, bm, xspace, pifont}
\usepackage{float}

\usepackage{hyperref}
\hypersetup{
 setpagesize=false, 
 colorlinks=true, 
 linkcolor=blue, 
 citecolor=blue, 
 urlcolor=blue
}

\begin{document}

\title{Light-Driven Intrinsic Perfect Superconducting Diode Effect}

\author{Makoto Ichikawa}
\email{ichikawa.makoto.45t@st.kyoto-u.ac.jp}

\author{Youichi Yanase}
\affiliation{Department of Physics, Graduate School of Science, Kyoto University, Kyoto 606-8502, Japan}

\date{\today}

\begin{abstract}
We demonstrate the perfect superconducting diode effect (SDE) --- unidirectional supercurrent with 100\% diode efficiency --- in light-driven nonequilibrium systems. 
Although the perfect SDE is difficult to achieve in equilibrium, monochromatic light induces the perfect SDE in systems lacking inversion and time-reversal symmetries. 
More strikingly, multi-frequency light enables the perfect SDE even in centrosymmetric systems via dynamical symmetry breaking. 
Our results establish a general principle for realizing unidirectional superconducting transport based on nonequilibrium control and symmetry engineering. 
\end{abstract}

\maketitle

\textit{Introduction}. --- 
Symmetry breaking is a central concept in condensed matter physics, since it often gives rise to unconventional properties. 
Recently, nonreciprocal transport arising from inversion and time-reversal symmetry breaking~\cite{Nagaosa2024} has attracted considerable attention. 
A prominent example is the superconducting diode effect (SDE)~\cite{Ando2020, Lyu2021, Mizuno2022, Narita2022, Bauriedl2022, Lin2022, Hou2023, Le2024, Qi2025, Nagata2025, Wu2022, Chiles2023, Valentini2024, Su2024, Wang2026, Daido2022, Yuan2022, He2022, Ilic2022, Daido2022-2, Ma2025, Hosur2023, Daido2025, Chakraborty2025, Daido2025-2, Li2026, Arora2026}, which represents the nonreciprocity of the critical current in superconductors. 
The SDE has been widely observed in various materials~\cite{Ando2020, Lyu2021, Mizuno2022, Narita2022, Bauriedl2022, Lin2022, Hou2023, Le2024, Qi2025, Nagata2025}, and in Josephson junctions, known as the Josephson diode effect~\cite{Wu2022, Chiles2023, Valentini2024, Su2024, Wang2026}. 
Thus, the SDE has potential for a sensitive probe of symmetry breaking in quantum materials and also for next-generation low-power devices. 

Recent interest in SDE has been triggered by its observation in Rashba-type Nb/V/Ta superlattices~\cite{Ando2020}. 
Following this report, extensive efforts have been devoted to elucidating intrinsic mechanisms of SDE originating from inversion symmetry breaking inherent in materials~\cite{Daido2022, Yuan2022, He2022, Ilic2022, Daido2022-2} and to exploring functionalities based on the SDE. 
The magnitude of SDE can be quantitatively evaluated by the diode efficiency $\eta$. 
Although $\eta$ remained at most a few percent in the initial study~\cite{Ando2020}, recent studies have demonstrated a significant enhancement of $\eta$~\cite{Ma2025, Wang2026}, approaching the ultimate goal of achieving the perfect SDE~\cite{Lin2022, Chiles2023, Valentini2024, Su2024, Wang2026, Hosur2023, Daido2025, Chakraborty2025, Daido2025-2}, a phenomenon in which the supercurrent can flow only in one direction. 
If an intrinsic perfect SDE is realized, the perfect SDE is expected to persist even in the presence of extrinsic mechanisms~\cite{Lyu2021, Hou2023} associated with vortex dynamics and so on, because extrinsic mechanisms require a finite current~\cite{Daido2025}. 
Therefore, the intrinsic perfect SDE will represent a milestone in the field of nonreciprocal transport in superconductors. 

In equilibrium systems, it is generally difficult to realize a perfect SDE~\cite{Daido2025}. 
Given this difficulty, extending the framework to nonequilibrium systems provides an effective route. 
Indeed, dissipation-induced perfect SDE has been proposed for nonequilibrium systems driven by a dc electric field~\cite{Daido2025}. 
In this context, attention naturally turns to nonequilibrium systems driven by ac electric fields under light or microwave irradiation. 
Notably, extensive studies have revealed the interplay between superconductivity and light-matter interactions, including light-induced superconductivity~\cite{Cavalleri2018}, observations of the Higgs mode~\cite{Matsunaga2013, Matsunaga2014}, superconducting switching phenomena~\cite{Sekiguchi2024}, and the microwave-irradiated diode effect~\cite{Valentini2024, Su2024, Wang2026, Arora2026}. 
In particular, the photocurrent arising from nonlinear optical responses~\cite{Watanabe2022, Tanaka2023, Tanaka2025, Matsyshyn2025} or a phase battery~\cite{Mironov2024} may modulate nonreciprocity of the dc supercurrent, potentially leading to unexplored SDEs. 
Despite these advances, light-driven intrinsic SDE remains largely unexplored. 

In this Letter, we extend the Ginzburg-Landau (GL) theory for the intrinsic SDE~\cite{Daido2022, Yuan2022, He2022} to nonequilibrium systems under light irradiation and investigate the light-driven intrinsic SDE. 
Using time-dependent GL (TDGL) equations, we analyze nonreciprocity of the dc component of the supercurrent. 
Our results reveal the following two key findings: (i) In systems lacking inversion and time-reversal symmetries, monochromatic light modulates the nonreciprocity and even induces the perfect SDE. 
Whereas the dc supercurrent arising from the second-order optical response provides the leading contribution for weak light intensity, higher even-order optical responses are indispensable for achieving the perfect SDE. 
(ii) Even in systems preserving inversion or time-reversal symmetry, multi-frequency light (e.g., $\omega$ and $2\omega$) induces the SDE via dynamical symmetry breaking. 
In this case, the dc supercurrent arising from odd-order optical responses contributes to nonreciprocity and even leads to the perfect SDE. 
As a result, even centrosymmetric materials, which exhibit reciprocal transport without light, can display nonreciprocal transport under light irradiation. 
These results indicate that combining the SDE with light-driven nonequilibrium superconductivity provides a promising platform for generating unconventional transport phenomena.

\textit{Model}. --- 
Hereafter, the units $\hbar=1$ and $e=1$ are adopted for convenience. 
TDGL equations are used to study the SDE under light irradiation near the critical temperature $T_{\mathrm{c}}$. 
We consider a one-dimensional wire or higher-dimensional systems with spatial modulation only along the $x$ axis. 
We adopt a single-$q$ ansatz for the order parameter $\Psi(x,t) = \psi(q,t)e^{iqx}$ with $\psi(q,t)\in\mathbb{R}$, where $q$ represents the center-of-mass momentum of Cooper pairs and is independent of time $t$, for simplicity. 

First, we consider the GL free energy density $f(q,\psi(q,t)) = \alpha(q)\psi(q,t)^2 + \frac{1}{2}\beta(q)\psi(q,t)^4$ in the absence of light. 
The GL coefficients $\alpha(q),\beta(q)$ are given by $\alpha(q) \coloneq \tilde{\alpha}_0 + \tilde{\alpha}_2(q-q_0)^2 + \tilde{\alpha}_3(q-q_0)^3$ and $\beta(q) \coloneq \tilde{\beta}_0 + \tilde{\beta}_1(q-q_0)$. 
This expansion has been microscopically derived in the Rashba-Zeeman model and is appropriate up to $O(h(T_{\mathrm{c}}-T)^{5/2})$~\cite{Daido2022, Daido2022-2} for the temperature $T$ and the magnetic field $h$ in the $xy$ plane. 
Replacing $q-q_0$ with $q$ and omitting the tildes for simplicity, we rewrite $\alpha(q) = \alpha_0 + \alpha_2q^2 + \alpha_3q^3,\,\beta(q) = \beta_0 + \beta_1q$. 
Odd-order terms $\alpha_3$ and $\beta_1$ are allowed in systems lacking inversion and time-reversal symmetries and contribute to nonreciprocity~\cite{Daido2022, Daido2022-2}. 

Next, we take into account ac electric fields arising from light or microwave irradiation. 
We consider continuous-wave linearly polarized light propagating along the $z$ direction, with the electric field $\vb*{E}(t)$ polarized along the $x$ direction. 
For monochromatic light, we take $\vb*{E}(t) = E_0\sin\omega t\,\hat{\vb*{x}}$, where $\hat{\vb*{x}}$ is the unit vector in the $x$ direction. 
Then, the vector potential is given by $\vb*{A}(t) = A_0\cos\omega t \,\hat{\vb*{x}} = A(t)\hat{\vb*{x}}$ with $A_0 = \frac{E_0}{\omega}$. 
The effect of light is incorporated by replacing $q$ in the GL coefficients with $q+2A(t)$. 
This corresponds to the replacement of the covariant derivative: $-i\nabla \rightarrow -i\nabla+2\vb*{A}(t)$. 
Accordingly, the GL free energy density in the presence of light is
\begin{align}
  \begin{split}
    f(q+2A(t),\psi(q,t))
    &= \alpha(q+2A(t))\psi(q,t)^2\\
    &\quad + \dfrac{1}{2}\beta(q+2A(t))\psi(q,t)^4.
  \end{split}
\end{align}

We assume that the time evolution of the order parameter is governed by TDGL equations with a first-order time derivative. 
Although its microscopic derivation is subject to certain restrictions~\cite{Cyrot1973, Kopnin2001, Tinkham2004}, it is an effective framework to study the SDE~\cite{Lyu2021, Daido2025, Li2026} and the interplay between superconductivity and light~\cite{Mironov2021, Gassner2024}. 
Using the free energy functional $F[\Psi,\Psi^{\ast},A]$ and its functional derivative, we write the TDGL equation as $\Gamma\pdv{\Psi(x,t)}{t} = -\frac{\delta F[\Psi,\Psi^{\ast},A]}{\delta\Psi^{\ast}}$, which leads to
\begin{equation}
  \label{Eq:TDGL}
  \Gamma\pdv{\psi(q,t)}{t} 
  = -[\alpha(q+2A(t))\psi(q,t) + \beta(q+2A(t))\psi(q,t)^3]. 
\end{equation}
Because of $\frac{\Im\,\Gamma}{\Re\,\Gamma} = O\left(T_{\mathrm{c}}/\varepsilon_{\mathrm{F}}\right)$~\cite{Ebisawa1971}, where $\varepsilon_{\mathrm{F}}$ is the Fermi energy, we neglect the imaginary part of $\Gamma$ and set $\Gamma>0$. 
The TDGL equation takes the form of a Bernoulli differential equation, and the solution is obtained as 
\begin{align}
  \begin{split}
    \label{Eq:Bernoulli}
    \dfrac{1}{\psi(q,t)^2}
    &=\Bigg[
        \dfrac{1}{\psi(q,0)^2}
        +\int_{0}^{t}\dd{t'}\,\dfrac{2\beta(q+2A(t'))}{\Gamma}\,
        \\
        &\times e^{-\int_{0}^{t'}\dd{t''}\,\frac{2\alpha(q+2A(t''))}{\Gamma}}
      \Bigg]
      e^{+\int_{0}^{t}\dd{t'}\,\frac{2\alpha(q+2A(t'))}{\Gamma}}.
  \end{split}
\end{align}
We focus on the long-time limit $t \rightarrow \infty$, in which the dependence on the initial state can be neglected. 
In the absence of light, Eq.~\eqref{Eq:Bernoulli} can be solved analytically as $\psi(q,t)^2 = -\frac{\alpha(q)}{\beta(q)}\frac{1}{1-C\exp(2\alpha(q)t/\Gamma)}$ with $C = \frac{\alpha(q)+\beta(q)\psi(q,0)^2}{\beta(q)\psi(q,0)^2}$. 
In the limit $t \rightarrow \infty$, $\psi(q,t)^2 \rightarrow \psi_{\mathrm{eq}}(q)^2 = -\frac{\alpha(q)}{\beta(q)}\theta(-\alpha(q))$, where $\theta(x)$ is the Heaviside step function. 
As expected, the solution reduces to the equilibrium state. 
Consequently, the free energy density reproduces the equilibrium value $f(q+2\cdot 0,\psi(q,t)) \rightarrow f_{\mathrm{eq}}(q) = f(q)\cdot\theta(-\alpha(q))$ with $f(q) \coloneq -\frac{\alpha(q)^2}{2\beta(q)}$, and the supercurrent defined below satisfies $j(q,t) \rightarrow j_{\mathrm{eq}}(q) = -2\dv{f_{\mathrm{eq}}(q)}{q}$. 

The time-dependent supercurrent is given by $j(q,t) = -\frac{\delta F[\Psi,\Psi^{\ast},A]}{\delta A}$, which leads to 
\begin{align}
  \begin{split}
    j(q,t)
    &= -2\Bigg[
      \pdv{\alpha(q+2A(t))}{q}\psi(q,t)^2\\
      &\qquad\qquad +\dfrac{1}{2}\pdv{\beta(q+2A(t))}{q}\psi(q,t)^4
      \Bigg]. \\ 
  \end{split}
\end{align} 
When the system is continuously irradiated by light with frequency $\omega$, the long-time behaviors of physical quantities become periodic in time, in analogy with Floquet theorem~\cite{Shirley1965, Oka2019}. 
Accordingly, the supercurrent can be expanded in a Fourier series, 
\begin{equation}
    j(q,t)
    = j_{\mathrm{dc}}(q)
      +\sum_{n=1}^{\infty}
      [
        j_{\cos, n}(q)\cos n\omega t
        +
        j_{\sin, n}(q)\sin n\omega t
      ]. 
\end{equation}
As we show below, superconductivity is suppressed, and the supercurrent vanishes when $\gamma_0(q) \coloneq \ev*{\alpha(q+2A(t))} > 0$. 
Here, $\ev*{\cdots}$ represents the time average over one period. 
The relation 
\begin{equation}
  \label{Eq:gamma_0}
  \ev*{\alpha(q+2A(t))}
  = \alpha(q) + \frac{\alpha''(q)}{4} \left(\frac{2E_0}{\omega}\right)^2,
\end{equation}
reveals that light suppresses superconductivity for $\alpha''(q)>0$. 

To demonstrate light-driven SDE, we focus on the dc supercurrent and decompose it as 
\begin{equation}
  j_{\mathrm{dc}}(q)
  = j_{\mathrm{eq'}}(q)
  + j_{\mathrm{photo}}(q), 
\end{equation}
where $j_{\mathrm{eq'}}(q) \coloneq j_{\mathrm{eq}}(q)\cdot\theta(-\gamma_0(q))$. 
The additional term $j_{\mathrm{photo}}(q)$ represents the modulation induced by light. 
The SDE is quantified by the diode efficiency $\eta$, defined as 
\begin{equation}
  \eta
  \coloneq
  \dfrac{j_{\mathrm{dc,c}+}-j_{\mathrm{dc,c}-}}{j_{\mathrm{dc,c}+}+j_{\mathrm{dc,c}-}}, 
\end{equation}
with $j_{\mathrm{dc,c}+} \coloneq \max_{q}j_{\mathrm{dc}}(q)$ and $j_{\mathrm{dc,c}-} \coloneq -\min_{q}j_{\mathrm{dc}}(q)$. 
Both $j_{\mathrm{dc,c} \pm}$ are non-negative in the usual setup, and the maximum efficiency is $|\eta|=1$. 
Reciprocal and nonreciprocal critical current correspond to $\eta=0$ and $\eta\neq 0$, respectively. 
With this definition, $|\eta|=1$ indicates the perfect SDE. 

In the following, we analyze the dimensionless form of the TDGL equation by setting $\Gamma=1$, $\alpha_0=-1$, $\alpha_2=1$, and $\beta_0=1$. 
Thus, the remaining parameters in the GL model are $\alpha_3$ and $\beta_1$.

\begin{figure}[tbp]
  \includegraphics[width=0.98\columnwidth]{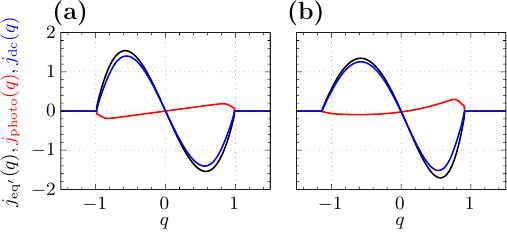}
  \caption{
    The dc supercurrent $j_{\mathrm{dc}}(q)$ (blue line) under light irradiation. 
    The equilibrium value $j_{\mathrm{eq'}}(q)$ (black line) is modulated by the photo-induced current $j_{\mathrm{photo}}(q)$ (red line). 
    The diode efficiency is (a) $\eta=0$ for $\alpha_3=0$ and (b) $\eta\sim-0.094$ for $\alpha_3=0.2$. 
    We set $\beta_1=0$, $E_0=0.01$, and $\omega=0.1$. 
  }
  \label{Fig1}
\end{figure}

\textit{SDE under monochromatic light}. ---
First, we discuss systems that preserve inversion or time-reversal symmetry, where $\alpha_3=0$ and $\beta_1=0$. 
Figure~\ref{Fig1}(a) shows $j_{\mathrm{dc}}(q)$ for a small monochromatic electric field $E_0$, i.e., $E_0 \ll \omega\sqrt{-\frac{\alpha_0}{\alpha_2}}$. 
In systems with these symmetries, the dc supercurrent satisfies a reciprocal relation $j_{\mathrm{dc}}(-q) = -j_{\mathrm{dc}}(q)$, which prohibits the SDE. 
The leading term of the photo-induced contribution $j_{\mathrm{photo}}(q)$ is proportional to ${E_0}^2$, corresponding to the second-order optical response. 
Even-order optical responses are forbidden by inversion symmetry. 
Indeed, at $q=0$, the most stable state in the absence of light, no supercurrent flows as symmetry enforces $j_{\mathrm{photo}}(0)=0$. 
In contrast, the inversion and time-reversal symmetries are broken when the momentum of Cooper pairs is finite $q\neq 0$, allowing supercurrent-induced even-order optical responses~\cite{Vaswani2020, Nakamura2020}. 
Symmetry enforces the reciprocal relation $j_{\mathrm{photo}}(-q) = -j_{\mathrm{photo}}(q)$, and thus, the photo-induced current does not contribute to the SDE. 
The same discussion applies to the second harmonic generation as well: $j_{\cos/\sin, 2}(-q) = -j_{\cos/\sin, 2}(q)$. 

Next, we discuss systems lacking inversion and time-reversal symmetries. 
Figure~\ref{Fig1}(b) shows $j_{\mathrm{dc}}(q)$ for small $E_0$ with $\alpha_3=0.2$ and $\beta_1=0$. 
In this case, $j_{\mathrm{eq'}}(q)$ and $j_{\mathrm{photo}}(q)$ break the reciprocal relation, and thus $j_{\mathrm{dc}}(q)$ is generally nonreciprocal. 
Through the photo-induced contribution $j_{\mathrm{photo}}(q)$, the diode efficiency $\eta$ can be tuned by light irradiation. 

\begin{figure}[tbp]
  \includegraphics[width=0.98\columnwidth]{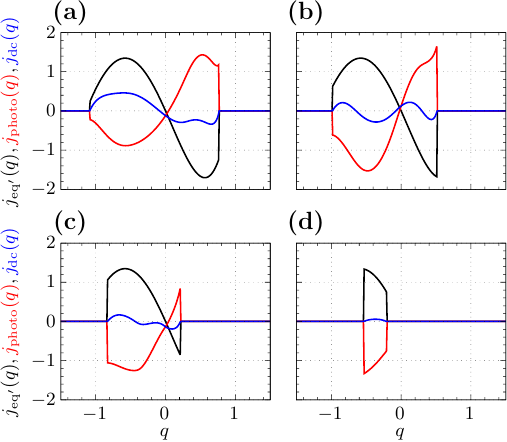}
  \caption{
    The dc supercurrent $j_{\mathrm{dc}}(q)$ and its components $j_{\mathrm{eq'}}(q)$ and $j_{\mathrm{photo}}(q)$ are shown by the same color as Fig.~\ref{Fig1}. 
    The ac electric field is increased from (a) to (d), as $E_0=0.033$, $0.052$, $0.065$, and $0.074$. The estimated diode efficiency is $\eta\sim 0.152$, $-0.124$, $-0.069$, and $1.000$, respectively. 
    The parameters are the same as in Fig.~\ref{Fig1}(b) except for $E_0$. 
  }
  \label{Fig2}
\end{figure}

\begin{figure}[tbp]
  \includegraphics[width=0.98\columnwidth]{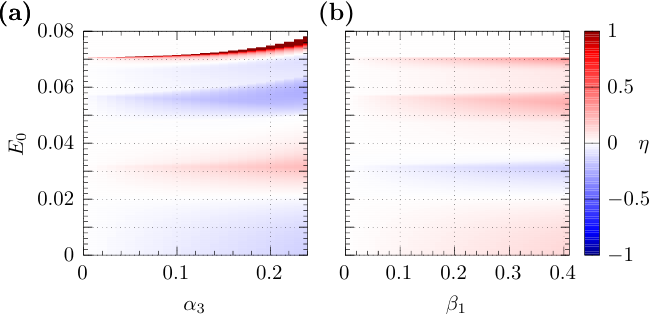}
  \caption{
    (a) Diode efficiency $\eta$ as a function of $\alpha_3$ and $E_0$ with $\beta_1=0$. 
    (b) $\beta_1$ and $E_0$ dependence of $\eta$ with $\alpha_3=0$. 
    We set $\omega=0.1$ in both (a) and (b). 
  }
  \label{Fig3}
\end{figure}

Figure~\ref{Fig2} shows the results for larger ac electric fields $E_0$ with fixed frequency $\omega$. 
As $E_0$ increases, superconductivity is suppressed as revealed in Eq.~\eqref{Eq:gamma_0}, and the dc supercurrent $j_{\mathrm{dc}}(q)$ shows complex behaviors. 
The momentum $q$ at which $j_{\mathrm{dc}}(q)$ is maximized or minimized changes discontinuously. 
Consequently, the diode efficiency non-monotonically changes as a function of light intensity (see Fig.~\ref{Fig3}). 
Interestingly, before superconductivity is completely suppressed, i.e., $\psi(q,t)\equiv 0$, the perfect SDE with ideal efficiency $\eta=1$ is realized as demonstrated in Fig.~\ref{Fig2}(d). 

Here, we clarify the key factor for the light-driven SDE. 
For this purpose, we estimate the diode efficiency $\eta$ for $\beta_1=0$ [$\alpha_3=0$] in Fig.~\ref{Fig3}(a) [Fig.~\ref{Fig3}(b)]. 
Although the intrinsic SDE stems from both $\alpha_3$ and $\beta_1$~\cite{Daido2022}, whose contributions are comparable~\cite{Ilic2022}, the light-driven perfect SDE is governed by $\alpha_3$. 
Indeed, when $\alpha_3=0$, the perfect SDE does not occur as shown in Fig.~\ref{Fig3}(b). 
In contrast, the perfect SDE emerges around $E_0\sim 0.07$ for finite $\alpha_3\neq 0$ as demonstrated in Fig.~\ref{Fig3}(a). 
Moreover, we have confirmed that the perfect SDE is ubiquitously achieved for light frequencies smaller than the superconducting gap~\cite{Supplement}; when $\alpha_3\neq 0$, it occurs near the critical ac field at which the superconducting state is completely destroyed. 

Subsequently, we compare the numerical results with an approximate solution of the TDGL equation~\cite{Supplement}, which leads to the photo-induced supercurrent up to ${E_0}^2$ order, 
\begin{align}
  \begin{split}
    j_{\mathrm{photo}}(q)
    &=
    \Bigg\{
      \dfrac{2\alpha\alpha'''}{\beta}
      +\dfrac{2(\alpha'\beta-\alpha\beta')}{(\alpha^2+\tilde{\omega}^2)\beta^2}
        \Bigg[
          \alpha''(3\alpha^2+\tilde{\omega}^2)\\
          &\quad -\dfrac{3\alpha^2\beta'(\alpha'\beta-\alpha\beta')}{\beta^2}
        \Bigg]
    \Bigg\}\frac{{E_0}^2}{\omega^2}
    + O\left(\frac{{E_0}^4}{\omega^4}\right). 
  \end{split}
\end{align}
Here, we introduce $\tilde{\omega} \coloneq \frac{\Gamma\omega}{2}$, and the arguments $q$ of $\alpha(q)$ and $\beta(q)$ are omitted. 
It should be noted that in the limit $\omega\rightarrow 0$, the approximate solution reduces to $j_{\mathrm{photo}}(q)|_{\omega\rightarrow 0} = -\frac{\rho_{\mathrm{NRSF}}(q)}{2\omega^2}\cdot\frac{{E_0}^2}{2}\cdot\theta(-\gamma_0(q)) + O\left(\frac{{E_0}^4}{\omega^4}\right)$ in agreement with the formula of the nonlinear optical responses~\cite{Watanabe2022} and the nonreciprocal Meissner effect~\cite{Watanabe2022-2}, where the nonreciprocal superfluid density (NRSF) is introduced by $\rho_{\mathrm{NRSF}}(q) \coloneq {\partial_{A}}^3f_{\mathrm{eq}}(q+2A)|_{A=0}$. 
This is reasonable because the dynamics of the order parameter can be neglected in the limit $\omega\rightarrow 0$ and Refs.~\cite{Watanabe2022, Watanabe2022-2} assume a static order parameter. 
Importantly, the relation holds for all $q$, including supercurrent-carrying states. 
The deviation between the numerical and approximate solutions grows as the ac electric field $E_0$ increases~\cite{Supplement}, although they coincide for small $E_0$. 
Crucially, higher-order optical responses beyond the second order are essential for achieving the perfect SDE.

\textit{SDE under two-frequency light}. ---
While we have focused on monochromatic light so far, we now consider the SDE under two-frequency light, in particular for systems that preserve inversion or time-reversal symmetry, i.e., $\alpha_3=0$ and $\beta_1=0$. 
Nonreciprocal dc transport can arise under two-frequency light even in centrosymmetric systems. 
Photocurrent generation by two-frequency light has been studied in centrosymmetric systems~\cite{Atanasov1996, Ikeda2023}. 
A crucial difference from the monochromatic case is that a dc current can be generated through odd-order optical responses. 
For example, light with frequencies $\omega$ and $2\omega$ induces a dc current through the third-order optical response, characterized by the third-order optical conductivity $\sigma^{(3)}(0;\omega,\omega,-2\omega)$. 
In contrast, for monochromatic light, only even-order optical responses give rise to a dc current, but they are prohibited in centrosymmetric systems. 

\begin{figure}[tbp]
  \includegraphics[width=0.98\columnwidth]{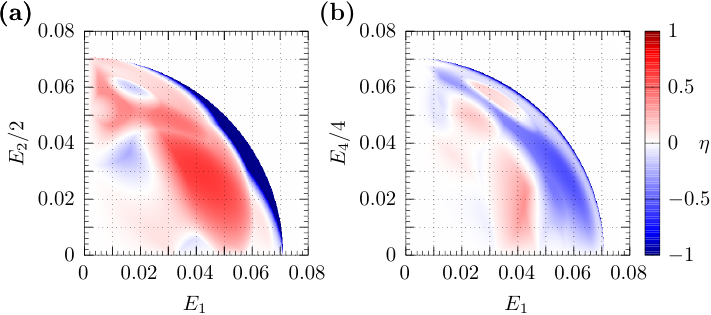}
  \caption{
    Diode efficiency $\eta$ in centrosymmetric systems ($\alpha_3=0$ and $\beta_1=0$) under two-frequency light. 
    The vector potential is $A(t) = A_1\cos\omega t + A_n\cos n\omega t$ with $\omega=0.1$. 
    (a) $n=2$. (b) $n=4$. 
  }
  \label{Fig4}
\end{figure}

Here, the TDGL equation is solved for two-frequency light by incorporating the vector potential $A(t) = A_1\cos\omega t + A_n\cos n\omega t$. 
Figure~\ref{Fig4} shows the diode efficiency $\eta$ for (a) $n=2$ and (b) $n=4$. 
Although the SDE is absent in equilibrium, the light-induced SDE occurs. 
Furthermore, as the ac electric fields $E_m =  m\omega A_m$ ($m=1,n$) increase, the perfect SDE with ideal efficiency $\eta=-1$ emerges near the critical ac electric field, as we have observed in noncentrosymmetric systems under monochromatic light. 

The condition for multi-frequency light that allows the light-induced SDE is as follows. 
The key feature is dynamical symmetry~\cite{Ben-Tal1993, Alon1998, Gassner2024}, which is represented as $A\left(t+\frac{T}{2}\right) = -A(t)$. 
Here, $T$ is the period of light. 
Dynamical symmetry is broken, namely $A(t+lT) \neq -A(t)$ for all $l\in(0,1)$, when two frequencies of light are not related by a ratio of two odd integers. 
Not only do even-multiple frequencies such as $\omega$ and $2\omega$ satisfy this condition, but more general cases such as $\omega$ and $1.5\omega$ do as well. 
Then, light-induced nonreciprocity is allowed. 

The essential role of dynamical symmetry breaking in the light-induced SDE is demonstrated as follows. 
Let us consider GL models that contain only $q$-even terms and assume dynamical symmetry of light $A\left(t+\frac{T}{2}\right) = -A(t)$. 
Then, considering the combined operation $q \rightarrow -q$ and $t \rightarrow t + \frac{T}{2}$ that transforms the order parameter as $\psi(q,t) \rightarrow \psi\left(-q,t+\frac{T}{2}\right)$, we find $\psi\left(-q,t+\tfrac{T}{2}\right)^2 = \psi(q,t)^2$, which follows from Eqs.~\eqref{Eq:TDGL} and \eqref{Eq:Bernoulli}. 
Here, dependence on the initial state is neglected in long-term behaviors $t \rightarrow \infty$. 
Therefore, we obtain 
\begin{align}
  \begin{split}   
    &f\left(-q+2A\left(t+\tfrac{T}{2}\right),\psi\left(-q,t+\tfrac{T}{2}\right)\right)
    \\
    &\qquad\qquad\qquad\qquad\qquad
    = f(q+2A(t),\psi(q,t)),
  \end{split}
\end{align}
which leads to
\begin{equation}
  \label{Eq:DS-j}
  j\left(-q,t+\tfrac{T}{2}\right) = -j(q,t). 
\end{equation}
Considering the temporal periodicity of $j(q,t)$ and taking the time average of Eq.~\eqref{Eq:DS-j}, we obtain $j_{\mathrm{dc}}(-q) = -j_{\mathrm{dc}}(q)$. 
Thus, dc transport is reciprocal in the presence of dynamical symmetry, and the SDE is prohibited. In contrast, breaking dynamical symmetry allows the light-induced SDE. 

We understand the above condition for the light-induced SDE based on symmetry of optical responses. 
When the GL coefficients contain only even-order terms in $q$, even-order optical responses are odd functions of $q$, while odd-order responses are even functions, even for multi-frequency light irradiation. 
Dynamical symmetry breaking allows odd-order optical responses that generate a dc current as an even function of $q$. 
This current distorts the reciprocal $q-j_{\mathrm{dc}}$ relation and results in the emergence of the SDE. 
From this point of view, we notice that the lowest-order contribution to nonreciprocity arises from the third-order optical response through $\sigma^{(3)}(0;\omega,\omega,-2\omega)$ for Fig.~\ref{Fig4}(a) while it arises from the fifth-order optical response $\sigma^{(5)}(0;\omega,\omega,\omega,\omega,-4\omega)$ for Fig.~\ref{Fig4}(b). 
Due to this difference, the diode efficiency is larger in Fig.~\ref{Fig4}(a) than in Fig.~\ref{Fig4}(b). 
The lowest-order contribution to the light-induced SDE is proportional to ${E_1}^2{E_2}$ and ${E_1}^4{E_4}$, respectively. 

For light with frequencies $\omega$ and $2\omega$, an approximate solution of the TDGL equation is obtained as described in Supplemental Material~\cite{Supplement}. 
A third-order photo-induced current $j_{\mathrm{photo}}^{(3)}(q)$ [Eq.~\eqref{Eq:j_{photo}^{(3)}} in Supplemental Material~\cite{Supplement}] is appropriately obtained in this approximate solution, and it is reduced to $j^{(3)}_{\mathrm{photo}}(q)|_{\omega\rightarrow 0} = -\frac{\rho_{\mathrm{NLSF}}(q)}{6\omega^2\cdot 2\omega}\cdot\frac{3{E_1}^2E_2}{4}\cdot\theta(-\gamma_0(q))$ for all $q$ in the limit $\omega\rightarrow 0$. This formula is regarded as a natural extension of the second-order photocurrent arising from the NRSF. 
Different from the NRSF, we define the nonlinear superfluid density (NLSF), $\rho_{\mathrm{NLSF}}(q) \coloneq {\partial_A}^4f_{\mathrm{eq}}(q+2A)|_{A=0}$. 
While the NRSF contributes to nonreciprocal transport in systems with broken inversion and time-reversal symmetries, the NLSF governs nonreciprocity through dynamical symmetry breaking. 
This is because the second-order optical response in superconductors is associated with the NRSF~\cite{Watanabe2022}, whereas the third-order optical response is related to its derivative, namely the NLSF.

\textit{Discussion}. ---
In this Letter, we demonstrated the light-driven and light-induced SDE which achieve perfect diode efficiency in the two cases: (i) systems lacking inversion and time-reversal symmetries under monochromatic light, and (ii) centrosymmetric materials where light breaks dynamical symmetry. 
Here, an external field of monochromatic light required to achieve the perfect SDE is estimated. 
We take the Fermi velocity $v_{\mathrm{F}}=10^6$ m/s and the transition temperature $T_{\mathrm{c}}=1$ K as typical parameters, and set the reduced temperature $T/T_{\mathrm{c}}=0.9$. 
Using the GL coefficients calculated from microscopic approaches~\cite{Kopnin2001, Ilic2022}, we estimate $E_0\sim 10^{-1}$ V/m for the perfect SDE that occurs at $\omega\sim 10^{9}$ rad/s (GHz range) corresponding to Fig.~\ref{Fig3}(a). 
Under the assumption of the antisymmetric spin-orbit coupling of $10^{-1}$ eV$\cdot\mathrm{\AA}$, the required magnetic field for $\alpha_3=10^{-1}$ in Fig.~\ref{Fig3}(a) is estimated to be $B\sim 10$ T. 
For high-temperature superconductors with $T_{\mathrm{c}}=10^2$ K and $v_{\mathrm{F}}=10^5$ m/s, we estimate $E_0\sim 10^4$ V/m at $\omega\sim 10^{11}$ rad/s, and $B\sim 10$ T for $\alpha_3=10^{-2}$ in Fig.~\ref{Fig3}(a). 
It should be noted that the required magnetic field can be reduced by employing heavy-electron systems~\cite{Stewart1984} with a small Fermi velocity or materials with a strong antisymmetric spin-orbit coupling~\cite{Ishizaka2011}. 

There remain several challenges in the realization of the light-driven perfect SDE. 
Since our analysis is based on the TDGL equation, it is restricted to temperatures slightly below $T_{\mathrm{c}}$. 
On the other hand, the present setup assumes continuous light irradiation, thus heating induced by light must be avoided. 
Studying microfabricated samples may be an effective way to suppress heating effects. 

Finally, we discuss the validity of the assumption that Cooper pair's momentum $q$ is independent of time. 
The characteristic time scale of the system is given by $\tau = \frac{\Gamma}{|\alpha_0|} =1$, whereas that of the external field is $\omega^{-1}$. 
Therefore, this assumption is justified for $\tau \gg \omega^{-1}$, that is, $\omega\gg 1$ in our unit, in which the order parameter cannot follow the oscillation of the external field. 
However, the order parameter represents the energy gap, and its magnitude is $O(1)$ in the present unit. 
Thus, the condition $\omega\gg 1$ is not consistent with the TDGL framework, which does not incorporate quasiparticle excitations. 
One possible way to maintain validity is to discretize $q$ by imposing periodic boundary conditions, as is accomplished in nanostructured superconductors. 
We emphasize that our TDGL analysis is intended as a minimal universal framework capturing symmetry-allowed nonreciprocal superconducting transport under periodic driving. 
The single-$q$ ansatz with time-independent $q$ simplifies the formulation and thereby allows us to obtain approximate solutions. 
Consequently, a transparent understanding of the light-driven SDE and nonlinear optical responses has been obtained.

\textit{Acknowledgments}. ---
We appreciate fruitful discussions with Akito Daido, Keito Hara, Yu Miura, Teruo Ono, and Hiroto Tanaka. 
This work is supported by JSPS KAKENHI (Grant Numbers JP22H01181, JP22H04933, JP23K17353, JP23K22452, JP24K21530, JP24H00007, JP25H01249, JP26H02016). 

\bibliography{ref}

\clearpage
\onecolumngrid
\input{supplemental_material}

\end{document}

%% file: supplemental_material.tex
\setcounter{equation}{0}
\setcounter{figure}{0}
\renewcommand{\theequation}{S\arabic{equation}}
\renewcommand{\thefigure}{S\arabic{figure}}

\begin{center}
{\bf \large Supplemental Material for\\ 
Light-Driven Intrinsic Perfect Superconducting Diode Effect}
\end{center}

\section{$\omega$-dependence of the diode efficiency $\eta$}

We show the $\omega$-dependence of the diode efficiency $\eta$. 
As can be seen in the time-dependent Ginzburg-Landau (TDGL) equation [Eq.~\eqref{Eq:TDGL} in the main text], for $l\in\mathbb{R}$, the scale transformation $\Gamma\rightarrow l\Gamma$ together with $t\rightarrow lt$ is equivalent to the transformation $E_0\rightarrow lE_0$ and $\omega\rightarrow l\omega$. 
Therefore, the following discussion can also be interpreted in terms of the $\Gamma$-dependence. 
However, their physical meanings are different. 
Whereas the dissipation constant $\Gamma$ is a material-dependent parameter, the electric field $E_0$ and the frequency $\omega$ are parameters of the applied light or microwave. 

\begin{figure}[H]
  \centering
  \includegraphics[width=0.4\columnwidth]{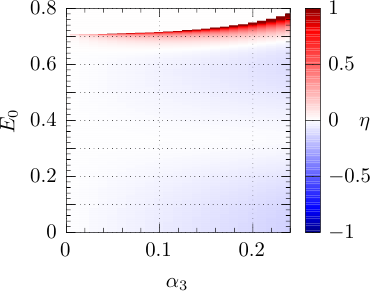}
  \caption{
    The diode efficiency $\eta$ as a function of $\alpha_3$ and $E_0$ for $\omega=1$. 
    Only $\omega$ is changed from Fig.~\ref{Fig3}(a) of the main text. 
  }
  \label{FigS1}
\end{figure}

Figure~\ref{FigS1} shows the result for $\omega=1$, which is comparable to the energy gap $\max_{q}|\psi_{\mathrm{eq}}(q)|$. 
Compared with Fig.~\ref{Fig3}(a) of the main text for $\omega=0.1$, only $\omega$ is changed while all other parameters are kept the same. 
Although the characteristic scale of the electric field $E_0$ is increased by one order of magnitude, a qualitative feature remains unchanged; the perfect SDE emerges in the large $E_0$ region. 
This suggests that the perfect SDE in the large $E_0$ region is a robust property for $\omega<\max_{q}|\psi_{\mathrm{eq}}(q)|$. 
However, in the small and intermediate $E_0$ region, the sign reversal of the diode efficiency $\eta$ is not observed in contrast to the result for $\omega=0.1$. 
This qualitative difference can be understood as follows. 
As $\omega$ increases, the order parameter $\psi(q,t)$ cannot follow the rapid oscillation of the electric field, resulting in a weaker dependence of $\eta$ on $E_0$. 
In the equivalent situation where $\Gamma$ increases, the time variation of $\psi(q,t)$ is suppressed, leading to the same conclusion. 

\section{Two-frequency light irradiation in nonreciprocal systems lacking symmetry}

\begin{figure}[H]
  \centering
  \includegraphics[width=0.8\columnwidth]{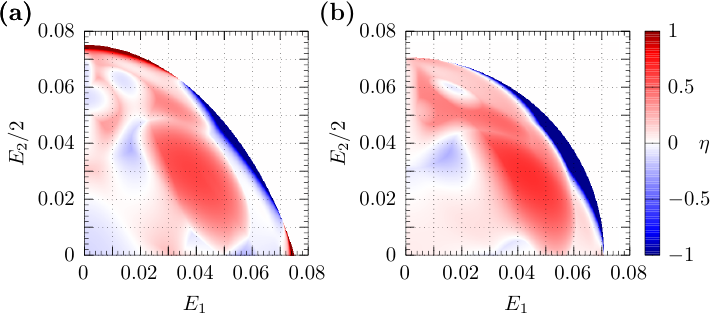}
  \caption{
    The diode efficiency $\eta$ for (a) $\alpha_3=0.2$ and $\beta_1=0$ and (b) $\alpha_3=0$ and $\beta_1=0.2$. 
    We consider the two-frequency light that gives rise to the vector potential $A(t)=A_1\cos\omega t+A_2\cos 2\omega t$ with $A_n\coloneq \frac{E_n}{n\omega}$ and $\omega=0.1$. 
  }
  \label{FigS2}
\end{figure}

We show the light-driven superconducting diode effect (SDE) in systems lacking inversion and time-reversal symmetries under two-frequency light irradiation. 
For $\alpha_3\neq 0$ or $\beta_1\neq 0$, we consider irradiation of light with frequencies $\omega$ and $2\omega$ ($\omega=0.1$). 
The results are compared with the case of monochromatic light irradiation in Fig.~\ref{Fig3} of the main text, and the case of two-frequency light for $\alpha_3=0$ and $\beta_1=0$ in Fig.~\ref{Fig4}(a) of the main text. 
We set the $x$-component of the vector potential $A(t) = A_1\cos\omega t + A_2\cos 2\omega t$, where $A_n\coloneq \frac{E_n}{n\omega}$. 
Figure~\ref{FigS2} shows the diode efficiency $\eta$. 
To distinguish the contributions of $\alpha_3$ and $\beta_1$, we consider the two cases: (a) $\alpha_3=0.2$, $\beta_1=0$, and (b) $\alpha_3=0$, $\beta_1=0.2$. 

On the lines $E_1=0$ or $E_2=0$, the results in Fig.~\ref{FigS2} are equivalent to those for monochromatic light. 
For the GL parameters in case (a), we see a perfect SDE with $\eta=1$ under monochromatic light, consistent with Fig.~\ref{Fig3}(a) in the main text. 
When the two Fourier components of the optical electric field are comparable in magnitude, a perfect SDE with an opposite sign $\eta=-1$ is realized, as shown in Fig.~\ref{FigS2}(a). 
The sign change of $\eta=\pm 1$ suggests that properties of light can dramatically modify nonreciprocity. 

In contrast, for the GL parameters in case (b), the results of SDE remain almost unchanged compared to Fig.~\ref{Fig4}(a) of the main text. 
This suggests that $\beta_1$ plays only a minor role in the realization of the perfect SDE even under two-frequency light irradiation. 

\section{Derivation of an approximate solution under monochromatic light irradiation}

We derive an approximate solution of the TDGL equation and the supercurrent under monochromatic light in a perturbative form. 
We set the $x$-component of the vector potential $A(t)=A_0\cos\omega t$. 
Hereafter, for notational simplicity, we redefine 
\begin{equation}
    A_0 \coloneq \frac{2E_0}{\omega}, 
\end{equation}
where the factor of $2$ in the argument of the GL coefficients $q+2\frac{E_0}{\omega}\cos\omega t$ is absorbed into the definition of $A_0$. 
It should be noted that this definition differs from that in the main text: $A_0 = \frac{E_0}{\omega}$. The final results do not depend on the choice of notation. 
From the TDGL equation, we obtain 
\begin{equation}
    \dfrac{1}{\psi(q,t)^2}
    =\left[
      \dfrac{1}{\psi(q,0)^2}
      +\int_{0}^{t}\dd{t'}\,
      \dfrac{2\beta(q+A(t'))}{\Gamma}\,
      e^{-\int_{0}^{t'}\dd{t''}\,\frac{2\alpha(q+A(t''))}{\Gamma}}
     \right]
     e^{+\int_{0}^{t}\dd{t'}\,\frac{2\alpha(q+A(t'))}{\Gamma}}. 
\end{equation}
We expand $\alpha(q+A(t))$ as 
\begin{align}
  \begin{split}
    \alpha(q+A(t))
    &=\alpha_0+\alpha_2(q+A(t))^2+\alpha_3(q+A(t))^3
    \\
    &=\alpha(q)+\alpha'(q)A(t)+\dfrac{1}{2}\alpha''(q)A(t)^2+\dfrac{1}{6}\alpha'''(q)A(t)^3
    \\
    &=\gamma_0(q)
     +\omega\gamma_1(q)\cos\omega t\cdot A_0
     +2\omega\gamma_2(q)\cos2\omega t\cdot {A_0}^2
     +3\omega\gamma_3(q)\cos3\omega t\cdot {A_0}^3. 
    \\
  \end{split}
\end{align}
Here, we introduce 
\begin{align}
  \begin{split}
    \gamma_0(q)=\alpha(q)+\dfrac{\alpha''(q)}{4}{A_0}^2=\ev{\alpha(q+A(t))},\quad
    \gamma_1(q)=\dfrac{\alpha'(q)+\dfrac{\alpha'''(q)}{8}{A_0}^2}{\omega},\quad
    \gamma_2(q)=\dfrac{\alpha''(q)}{8\omega},\quad
    \gamma_3(q)=\dfrac{\alpha'''(q)}{72\omega}, 
  \end{split}
\end{align}
where $\ev{\cdots}$ denotes the time average over one period. 
Similarly, we expand $\beta(q+A(t))$ as 
\begin{align}
  \begin{split}
    \beta(q+A(t))
    &=\beta_0+\beta_1(q+A(t))
    \\
    &=\beta(q)+\beta'(q)A(t)
    \\
    &=\beta(q)+\beta'(q)\cos\omega t\cdot A_0. 
    \\
  \end{split}
\end{align}
For notational simplicity, we omit the prefactor $\frac{2}{\Gamma}$ multiplying the GL coefficients for a while, restoring it if necessary. 
Then, 
\begin{equation}
  \exp(\int_{0}^{t}\dd{t'}\, \alpha(q+A(t')))
  =e^{\gamma_0(q) t}\cdot
  \exp(\sum_{n=1}^{3}\gamma_n(q) \sin n\omega t\cdot {A_0}^n). 
\end{equation}
For $q$ such that $\gamma_0(q)>0$, $e^{\gamma_0(q) t}\rightarrow\infty$ as $t\rightarrow\infty$, and for $q$ such that $\gamma_0(q)<0$, $e^{\gamma_0(q) t}\rightarrow 0$ as $t\rightarrow\infty$. 
In contrast, the factor involving $\gamma_n(q)$ ($n=1,2,3$) remains bounded, i.e., $\left|\exp(\sum_{n=1}^{3}\gamma_n(q) \sin n\omega t\cdot {A_0}^n)\right|<\infty$. 
Therefore, the long-term behavior in the limit $t\rightarrow\infty$ is governed mainly by $e^{\gamma_0(q) t}$ rather than the oscillatory term involving $\gamma_n(q)$. 
For $q$ such that $\gamma_0(q)>0$, the denominator of $\psi(q,t)$ diverges in the limit $t\rightarrow\infty$, which leads to $\psi(q,t)\rightarrow 0$. 
In contrast, for $q$ such that $\gamma_0(q)<0$, $\psi(q,t)$ remains finite in the limit  $t\rightarrow\infty$ and the dependence on the initial value becomes negligible. 
In the equilibrium state without light irradiation, the GL theory tells us that $\psi_{\mathrm{eq}}(q)=0$ for $q$ such that $\alpha(q)>0$, and $\psi_{\mathrm{eq}}(q)\neq 0$ for $q$ such that $\alpha(q)<0$. 
Under light irradiation, the role of $\alpha(q)$ is replaced by $\gamma_0(q)=\ev{\alpha(q+A(t))}$. 
In the following, we focus on $q$ such that $\gamma_0(q)<0$. Otherwise, superconductivity is suppressed in the limit $t\rightarrow\infty$. 

To obtain an approximate solution, we expand $\exp(-\sum_{n=1}^{3}\gamma_n(q) \sin n\omega t\cdot {A_0}^n)$ in powers of $A_0$. 
This expansion is valid when
\begin{equation}
  \left|\dfrac{\alpha^{(n)}(q){A_0}^n}{\Gamma\omega}\right| \ll 1,\quad(n=1,2,3), 
\end{equation}
where the omitted prefactor $\frac{2}{\Gamma}$ has been restored. 
We retain terms up to ${A_0}^3$ order, 
\begin{align}
  \begin{split}
    &\exp(-\int_{0}^{t}\dd{t'}\, \alpha(q+A(t')))
    \\
    =&\, e^{-\gamma_0(q) t}\cdot
      \exp(-\sum_{n=1}^{3}\gamma_n(q) \sin n\omega t\cdot {A_0}^n)
    \\
    \sim&\, e^{-\gamma_0(q) t}\cdot
      \Bigg[
        1-\sum_{n=1}^{3}\gamma_n(q) \sin n\omega t\cdot {A_0}^n
        \\
        &\qquad\qquad\qquad\qquad
        +\dfrac{1}{2}\left(
          {\gamma_1(q)}^2\sin^2\omega t\cdot {A_0}^2
          +2\gamma_1(q)\gamma_2(q)\sin\omega t\sin 2\omega t\cdot {A_0}^3
        \right)
        -\dfrac{1}{6}{\gamma_1(q)}^3\sin^3\omega t\cdot {A_0}^3
      \Bigg]
    \\
    \sim&\, e^{-\gamma_0(q) t}\cdot
      \Bigg[
        1-\gamma_{1,0}(q)\sin \omega t\cdot A_0
        +\left(
          -\gamma_2(q)\sin 2\omega t
          +\dfrac{1}{2}{\gamma_{1,0}(q)}^2\sin^2\omega t
        \right)\cdot{A_0}^2
        \\
        &\qquad\qquad\qquad\qquad
        +\left(
          -\gamma_{1,2}(q)\sin\omega t
          -\gamma_3(q)\sin 3\omega t
          +\gamma_{1,0}(q)\gamma_2(q)\sin\omega t\sin 2\omega t
          -\dfrac{1}{6}{\gamma_{1,0}(q)}^3\sin^3\omega t
        \right)\cdot {A_0}^3
      \Bigg]. 
  \end{split}
\end{align}
Here, we have introduced 
\begin{equation}
  \gamma_{1,0}(q)\coloneq\dfrac{\alpha'(q)}{\omega},\quad
  \gamma_{1,2}(q)\coloneq\dfrac{\alpha'''(q)}{8\omega}. 
\end{equation}
Then, we evaluate 
\begin{align}
  \begin{split}
    \label{Eq:Bernoulli-part}
    &\int_{0}^{t}\dd{t'}\,
      \beta(q+A(t'))\,
      e^{-\int_{0}^{t'}\dd{t''}\, \alpha(q+A(t''))}\\
    \sim &\int_{0}^{t}\dd{t'}\,
      e^{-\gamma_0(q) t'}\big(\beta(q)+\beta'(q)\cos\omega t'\cdot A_0\big)\times
       \Bigg[
        1-\gamma_{1,0}(q)\sin \omega t'\cdot A_0
        +\left(
          -\gamma_2(q)\sin 2\omega t'
          +\dfrac{1}{2}{\gamma_{1,0}(q)}^2\sin^2\omega t'
        \right)\cdot{A_0}^2
        \\
        &\qquad\qquad\qquad\qquad\qquad
        +\left(
          -\gamma_{1,2}(q)\sin\omega t'
          -\gamma_3(q)\sin 3\omega t'
          +\gamma_{1,0}(q)\gamma_2(q)\sin\omega t'\sin 2\omega t'
          -\dfrac{1}{6}{\gamma_{1,0}(q)}^3\sin^3\omega t'
        \right)\cdot {A_0}^3
      \Bigg], 
  \end{split}
\end{align}
using the following relations 
\begin{align}
  \begin{split}
    \int_{0}^{t}\dd{t'}\, e^{-\gamma_0(q) t'}\cos n\omega t'
    &=\dfrac{1}{{\gamma_0(q)}^2+(n\omega)^2}
      \left[
        e^{-\gamma_0(q) t}(-\gamma_0(q)\cos n\omega t + n\omega\sin n\omega t) + \gamma_0(q)
      \right], 
    \\
    \int_{0}^{t}\dd{t'}\, e^{-\gamma_0(q) t'}\sin n\omega t'
    &=\dfrac{1}{{\gamma_0(q)}^2+(n\omega)^2}
      \left[
        e^{-\gamma_0(q) t}(-\gamma_0(q)\sin n\omega t - n\omega\cos n\omega t) + n\omega
      \right]. 
    \\
  \end{split}
\end{align}
We multiply Eq.~\eqref{Eq:Bernoulli-part} by 
\begin{align}
  \begin{split}
    &\exp(+\int_{0}^{t}\dd{t'}\, \alpha(q+A(t')))
    \\
    \sim&\, e^{\gamma_0(q) t}\cdot
       \Bigg[
        1+\gamma_{1,0}(q)\sin \omega t\cdot A_0
        +\left(
          \gamma_2(q)\sin 2\omega t
          +\dfrac{1}{2}{\gamma_{1,0}(q)}^2\sin^2\omega t
        \right)\cdot{A_0}^2
        \\
        &\qquad\qquad\qquad\qquad
        +\left(
          \gamma_{1,2}(q)\sin\omega t
          +\gamma_3(q)\sin 3\omega t
          +\gamma_{1,0}(q)\gamma_2(q)\sin\omega t\sin 2\omega t
          +\dfrac{1}{6}{\gamma_{1,0}(q)}^3\sin^3\omega t
        \right)\cdot {A_0}^3
      \Bigg],
  \end{split}
\end{align}
and retain terms up to ${A_0}^3$ order. 
Because we are interested in the region of $q$ satisfying $\gamma_0(q)<0$, where $\psi(q,t)$ is nonzero in the limit $t\rightarrow\infty$, we can neglect terms involving $e^{\gamma_0(q) t}$ and regard $\psi(q,t)$ as the periodic steady-state solution. 

Performing the calculation, we find 
\begin{equation}
  \label{Eq:psi^{-2}}
  \dfrac{1}{\psi(q,t)^2}
  \sim C_0(q)+C_1(q,t)A_0+C_2(q,t){A_0}^2+C_3(q,t){A_0}^3. 
\end{equation}
Here, we obtain 
\begin{align}
  \begin{split}
    C_0(q) &= -\dfrac{\beta(q)}{\alpha(q)}, \\
    C_1(q,t) &= C_{1;\cos,1}(q)\cos\omega t +C_{1;\sin,1}(q)\sin\omega t, \\
    C_2(q,t) &= C_{2;\mathrm{dc}}(q)+C_{2;\cos,2}(q)\cos 2\omega t +C_{2;\sin,2}(q)\sin 2\omega t, \\
    C_3(q,t) &= C_{3;\cos,1}(q)\cos\omega t + C_{3;\sin,1}(q)\sin\omega t 
                + C_{3;\cos,3}(q)\cos 3\omega t + C_{3;\sin,3}(q)\sin 3\omega t, \\
  \end{split}
\end{align}
where 
\begin{equation}
  C_{1;\cos,1}(q)
  =
  \dfrac{\alpha'\beta-\alpha\beta'}{\alpha^2+\tilde{\omega}^2}, 
\end{equation}
\begin{equation}
  C_{1;\sin,1}(q)
  =
  -\dfrac{\tilde{\omega}}{\alpha}
  \dfrac{\alpha'\beta-\alpha\beta'}{\alpha^2+\tilde{\omega}^2}
  =
  -\dfrac{\tilde{\omega}}{\alpha}C_{1;\cos,1}(q), 
\end{equation}
\begin{equation}
  C_{2;\mathrm{dc}}(q)
  =
  \dfrac{\alpha''\beta}{4\alpha^2}
  -\dfrac{\dfrac{\alpha'}{2\alpha}\big(\alpha'\beta-\alpha\beta'\big)}{\alpha^2+\tilde{\omega}^2}, 
\end{equation}
\begin{align}
  \begin{split}
    C_{2;\cos,2}(q)
    =
    \dfrac{\dfrac{\alpha''\beta}{4}+\alpha'\beta'+\dfrac{\alpha\alpha'^2\beta}{4\tilde{\omega}^2}}{\alpha^2+4\tilde{\omega}^2}
    +\dfrac{\dfrac{\alpha'}{4\alpha}\big(\alpha'\beta-\alpha\beta'\big)-\dfrac{\alpha'\beta'}{4}-\dfrac{\alpha\alpha'^2\beta}{4\tilde{\omega}^2}}{\alpha^2+\tilde{\omega}^2}, 
  \end{split}
\end{align}
\begin{align}
  \begin{split}
    C_{2;\sin,2}(q)
    =
    -\dfrac{\dfrac{\alpha''\beta\tilde{\omega}}{2\alpha}+\dfrac{\alpha'}{2\tilde{\omega}}\big(\alpha'\beta-\alpha\beta'\big)}{\alpha^2+4\tilde{\omega}^2}
    +\dfrac{\dfrac{\alpha'}{2\tilde{\omega}}\big(\alpha'\beta-\alpha\beta'\big)}{\alpha^2+\tilde{\omega}^2}, 
  \end{split}
\end{align}
\begin{align}
  \begin{split}
    C_{3;\cos,1}(q)
    =
    -\dfrac{\dfrac{\alpha'\alpha''\beta}{4\alpha}+\dfrac{\alpha'^2}{4\tilde{\omega}^2}\big(\alpha'\beta-\alpha\beta'\big)}{\alpha^2+4\tilde{\omega}^2}
    +\dfrac{\dfrac{\alpha'''\beta}{8}-\dfrac{\alpha''\beta'}{8}+\dfrac{\alpha'^2}{4\tilde{\omega}^2}\big(\alpha'\beta-\alpha\beta'\big)}{\alpha^2+\tilde{\omega}^2}
    -\dfrac{\dfrac{\alpha\alpha''}{2}\big(\alpha'\beta-\alpha\beta'\big)}{\big(\alpha^2+\tilde{\omega}^2\big)^2}, 
  \end{split}
\end{align}
\begin{align}
  \begin{split}
    C_{3;\sin,1}(q)
    =
    \dfrac{\dfrac{\alpha'\alpha''\beta\tilde{\omega}}{2\alpha^2}+\dfrac{\alpha'^2}{2\alpha\tilde{\omega}}\big(\alpha'\beta-\alpha\beta'\big)}{\alpha^2+4\tilde{\omega}^2}
    +\dfrac{-\dfrac{\alpha'''\beta\tilde{\omega}}{8\alpha}+\dfrac{\alpha'\alpha''\beta\tilde{\omega}}{8\alpha^2}-\dfrac{\alpha'^2}{2\alpha\tilde{\omega}}\big(\alpha'\beta-\alpha\beta'\big)}{\alpha^2+\tilde{\omega}^2}
    +\dfrac{\dfrac{\alpha''\tilde{\omega}}{2}\big(\alpha'\beta-\alpha\beta'\big)}{\big(\alpha^2+\tilde{\omega}^2\big)^2}, 
  \end{split}
\end{align}
\begin{align}
  \begin{split}
    C_{3;\cos,3}(q)
    &=
    \dfrac{\dfrac{\alpha'''\beta}{24}+\dfrac{3\alpha''\beta'}{16}+\dfrac{\alpha\alpha'\alpha''\beta}{16\tilde{\omega}^2}-\dfrac{\alpha'^2}{8\tilde{\omega}^2}\big(\alpha'\beta-\alpha\beta'\big)}{\alpha^2+9\tilde{\omega}^2}
    +\dfrac{-\dfrac{\alpha\alpha'\alpha''\beta}{16\tilde{\omega}^2}+\dfrac{\alpha'^2}{4\tilde{\omega}^2}\big(\alpha'\beta-\alpha\beta'\big)}{\alpha^2+4\tilde{\omega}^2}
    \\
    &\qquad
    +\dfrac{\left(\dfrac{\alpha''}{16\alpha}-\dfrac{\alpha'^2}{8\tilde{\omega}^2}\right)\big(\alpha'\beta-\alpha\beta'\big)}{\alpha^2+\tilde{\omega}^2}, 
  \end{split}
\end{align}
\begin{align}
  \begin{split}
    C_{3;\sin,3}(q)
    &=
    \dfrac{-\dfrac{\alpha'''\beta\tilde{\omega}}{8\alpha}-\dfrac{\alpha'\alpha''\beta}{8\tilde{\omega}}+\left(-\dfrac{\alpha''}{16\tilde{\omega}}+\dfrac{3\alpha'^2}{8\alpha\tilde{\omega}}\right)\big(\alpha'\beta-\alpha\beta'\big)}{\alpha^2+9\tilde{\omega}^2}
    +\dfrac{\dfrac{\alpha'\alpha''\beta}{8\tilde{\omega}}-\dfrac{\alpha'^2}{2\alpha\tilde{\omega}}\big(\alpha'\beta-\alpha\beta'\big)}{\alpha^2+4\tilde{\omega}^2}
    \\
    &\qquad
    +\dfrac{\left(\dfrac{\alpha''}{16\tilde{\omega}}+\dfrac{\alpha'^2}{8\alpha\tilde{\omega}}\right)\big(\alpha'\beta-\alpha\beta'\big)}{\alpha^2+\tilde{\omega}^2}.  
  \end{split}
\end{align}
Here, we omit the argument $q$ of the GL coefficients $\alpha(q)$ and $\beta(q)$ and define 
\begin{equation}
  \tilde{\omega}\coloneq \dfrac{\Gamma\omega}{2}, 
\end{equation}
which incorporates the omitted prefactor $\frac{2}{\Gamma}$ into $\omega$. 
In the above derivation, we used the relations 
\begin{equation}
  \dfrac{1}{\gamma_0(q)}
  \sim
  \dfrac{1}{\alpha(q)}
  \left(
    1-\dfrac{\alpha''(q)}{4\alpha(q)}{A_0}^2
  \right),\quad
  \dfrac{1}{\gamma_0(q)^2+\tilde{\omega}^2}
  \sim
  \dfrac{1}{\alpha(q)^2+\tilde{\omega}^2}
  \left[
    1-\dfrac{\alpha(q)\alpha''(q)}{2\big(\alpha(q)^2+\tilde{\omega}^2\big)}{A_0}^2
  \right]. 
\end{equation}
Therefore, this approximation is valid when 
\begin{equation}
  \left|\dfrac{\alpha''(q)}{\alpha(q)}{A_0}^2\right|\ll 1,\quad
  \left|\dfrac{\alpha(q)\alpha''(q)}{\alpha(q)^2+\tilde{\omega}^2}{A_0}^2\right|\ll 1, 
\end{equation}
which indicates that the approximation breaks down as $|\alpha(q)|\rightarrow 0$. 

Using Eq.~\eqref{Eq:psi^{-2}}, we obtain an approximate solution of the TDGL equation, 
\begin{align}
  \begin{split}
    \label{Eq:psi^2}
    \psi(q,t)^2
    &\sim
     \dfrac{1}{C_0(q)}
     \left[
      1-\dfrac{C_1(q,t)}{C_0(q)}A_0
      +\left(
        \dfrac{{C_1(q,t)}^2}{{C_0(q)}^2}
        -\dfrac{C_2(q,t)}{C_0(q)}
       \right){A_0}^2
      +\left(
        -\dfrac{{C_1(q,t)}^3}{{C_0(q)}^3}
        +\dfrac{2C_1(q,t)C_2(q,t)}{{C_0(q)}^2}
        -\dfrac{C_3(q,t)}{C_0(q)}        
       \right){A_0}^3
     \right]
     \\
    &=\dfrac{1}{C_0(q)}
      \Big(1+\tilde{C}_1(q,t)A_0
        +\tilde{C}_2(q,t){A_0}^2
        +\tilde{C}_3(q,t){A_0}^3\Big). 
  \end{split}
\end{align}
Substituting this expression into the formula of supercurrent $j(q,t)$ yields the expression up to ${A_0}^3$ order. 
\begin{align}
  \begin{split}
    j(q,t)
    &=
    -2\left(
      \pdv{\alpha(q+A(t))}{q}\psi(q,t)^2
      +\dfrac{1}{2}\pdv{\beta(q+A(t))}{q}\psi(q,t)^4
    \right)\\
    &\sim
    -\dfrac{2}{C_0(q)}
      \Bigg[
        \alpha'(q)
        +\Big(\alpha''(q)\cos\omega t +\alpha'(q)\cdot\tilde{C}_1(q,t)\Big)A_0
        \\
        &\qquad
        +\left(
          \dfrac{\alpha'''(q)}{2}\cos^2\omega t+\alpha''(q)\cos\omega t\cdot\tilde{C}_1(q,t) + \alpha'(q)\cdot\tilde{C}_2(q,t)
        \right){A_0}^2
        \\
        &\qquad 
        +\left(
          \dfrac{\alpha'''(q)}{2}\cos^2\omega t\cdot\tilde{C}_1(q,t)+\alpha''(q)\cos\omega t\cdot\tilde{C}_2(q,t) + \alpha'(q)\cdot\tilde{C}_3(q,t)
        \right){A_0}^3
      \Bigg]\\
    &\quad 
    -\dfrac{\beta'(q)}{{C_0(q)}^2}
      \left[
        1
        +2\tilde{C}_1(q,t)A_0
        +\left({\tilde{C}_1(q,t)}^2+2\tilde{C}_2(q,t)\right){A_0}^2
        +2\Big(\tilde{C}_1(q,t)\tilde{C}_2(q,t)+\tilde{C}_3(q,t)\Big){A_0}^3
      \right]
    \\
    &=
    j_{\mathrm{eq'}}(q)
    +
    \left[
      \dfrac{2\alpha\alpha''}{\beta}\cos\omega t
      +
      \dfrac{2\alpha^2\big(\alpha'\beta-\alpha\beta'\big)}{\beta^3}C_1(q,t)
    \right]A_0
    \\
    &\quad
    +
    \left[
      \dfrac{\alpha\alpha'''}{\beta}\cos^2\omega t
      +
      \dfrac{2\alpha^2\alpha''}{\beta^2}\cos\omega t \cdot C_1(q,t)
      +
      \dfrac{\alpha^3\big(2\alpha'\beta-3\alpha\beta'\big)}{\beta^4}C_1(q,t)^2
      +
      \dfrac{2\alpha^2\big(\alpha'\beta-\alpha\beta'\big)}{\beta^3}C_2(q,t)
    \right]{A_0}^2
    \\
    &\quad
    +
    \Bigg[
      \dfrac{\alpha^2\alpha'''}{\beta^2}\cos^2\omega t \cdot C_1(q,t)
      +
      \dfrac{2\alpha^3\alpha''}{\beta^3}\cos\omega t \cdot C_1(q,t)^2
      +
      \dfrac{2\alpha^2\alpha''}{\beta^2}\cos\omega t \cdot C_2(q,t)
    \\
    &\quad\quad
    +
    \dfrac{2\alpha^4\big(\alpha'\beta-2\alpha\beta'\big)}{\beta^5} C_1(q,t)^3
    +
    \dfrac{2\alpha^3\big(2\alpha'\beta-3\alpha\beta'\big)}{\beta^4} C_1(q,t)C_2(q,t)
    +
    \dfrac{2\alpha^2\big(\alpha'\beta-\alpha\beta'\big)}{\beta^3}C_3(q,t)
    \Bigg]{A_0}^3. 
  \end{split}
\end{align}
Here, we again omit the argument $q$ of the GL coefficients $\alpha(q)$ and $\beta(q)$. 

As discussed above, the order parameter $\psi(q,t)$ becomes periodic for time with period $T=\frac{2\pi}{\omega}$. 
Consequently, the supercurrent $j(q,t)$ also becomes periodic and can be expanded in a Fourier series: 
\begin{align}
  \begin{split}
    j(q,t)
    &=j_{\mathrm{dc}}(q)
    +\sum_{n=1}^{\infty}
      \left[
        j_{\cos,n}(q)\cos n\omega t
        + j_{\sin,n}(q)\sin n\omega t
      \right], 
    \\
    j_{\mathrm{dc}}(q)
    &=j_{\mathrm{eq'}}(q) + j_{\mathrm{photo}}(q), 
    \\
  \end{split}
\end{align}
where the expansion coefficients are given by 
\begin{align}
  \begin{split}
  \label{Eq:j_coefficients}
    j_{\mathrm{dc}}(q)
    &=\dfrac{1}{T}\int_{NT}^{(N+1)T}\dd{t}\, j(q,t), 
    \\
    j_{\cos,n}(q)
    &=\dfrac{2}{T}\int_{NT}^{(N+1)T}\dd{t}\, j(q,t)\cos n\omega t, 
    \\
    j_{\sin,n}(q)
    &=\dfrac{2}{T}\int_{NT}^{(N+1)T}\dd{t}\, j(q,t)\sin n\omega t. 
    \\
  \end{split}
\end{align}
Here, $N\in\mathbb{N}$, and the integral is taken over one period after the system has reached the periodic regime. 
The coefficients are expanded in powers of $A_0$: 
\begin{align}
  \begin{split}
  \label{Eq:expansion_of_coefficients}
    j_{\mathrm{photo}}(q)
    &=\sum_{n=1}^{\infty}j_{\mathrm{photo}}^{(2n)}(q),
    \quad j_{\mathrm{photo}}^{(2n)}(q)\propto{A_0}^{2n}, \\
    j_{\cos,n}(q)
    &=\sum_{m=0}^{\infty}j_{\cos,n}^{(n+2m)}(q),
    \quad j_{\cos,n}^{(n+2m)}(q)\propto{A_0}^{n+2m}, \\
    j_{\sin,n}(q)
    &=\sum_{m=0}^{\infty}j_{\sin,n}^{(n+2m)}(q),
    \quad j_{\sin,n}^{(n+2m)}(q)\propto{A_0}^{n+2m}, \\
  \end{split}
\end{align}
in the monochromatic case. 

When the applied ac electric field is monochromatic, dynamical symmetry of light is preserved. If the GL model contains only $q$-even terms, 
the dynamical symmetry leads to $j\left(-q,t+\tfrac{T}{2}\right)=-j(q,t)$. 
Thus, from Eq.~\eqref{Eq:j_coefficients}, we obtain $j_{\mathrm{dc}}(-q)=-j_{\mathrm{dc}}(q)$ and 
\begin{equation}
  j_{\cos,n}(-q)=(-1)^{n-1}j_{\cos,n}(q),\quad
  j_{\sin,n}(-q)=(-1)^{n-1}j_{\sin,n}(q). 
\end{equation}
Therefore, even-order optical responses (including the dc photocurrent response) are odd functions of $q$, for example $j_{\cos/\sin,2}(-q)=-j_{\cos/\sin,2}(q)$, whereas odd-order optical responses are even functions of $q$. 
However, these relations are broken in the presence of odd-order terms of $q$ in the GL coefficients, $\alpha(q)$ and $\beta(q)$, which are allowed when inversion and time-reversal symmetries are broken. 

Here, we summarize the results of the calculations for Eq.~\eqref{Eq:expansion_of_coefficients}, where the argument $q$ of the GL coefficients $\alpha(q)$ and $\beta(q)$ are omitted. 
\begin{align}
  \begin{split}
    \dfrac{j_{\mathrm{photo}}^{(2)}(q)}{{A_0}^2}
    &=
    \dfrac{\alpha\alpha'''}{2\beta}
    +
    \dfrac{2\alpha^2\alpha''}{\beta^2}\dfrac{C_{1;\cos,1}(q)}{2}
    +
    \dfrac{\alpha^3\big(2\alpha'\beta-3\alpha\beta'\big)}{\beta^4}\dfrac{C_{1;\cos,1}(q)^2+C_{1;\sin,1}(q)^2}{2}
    +
    \dfrac{2\alpha^2\big(\alpha'\beta-\alpha\beta'\big)}{\beta^3}C_{2;\mathrm{dc}}(q)
    \\
    &=
    \dfrac{\alpha\alpha'''}{2\beta}
    +
    \dfrac{\alpha'\beta-\alpha\beta'}{2\left(\alpha^2+{\tilde{\omega}}^2\right)\beta^2}
      \left[
        \alpha''\left(
          3\alpha^2+{\tilde{\omega}}^2
        \right)
        -\dfrac{3\alpha^2\beta'(\alpha'\beta-\alpha\beta')}{\beta^2}
      \right], 
  \end{split}
\end{align}
\begin{align}
  \begin{split}
    \dfrac{j_{\cos,1}^{(1)}(q)}{A_0}
    &=
    \dfrac{2\alpha\alpha''}{\beta}
    +
    \dfrac{2\alpha^2\left(\alpha'\beta-\alpha\beta'\right)}{\beta^3}C_{1;\cos,1}(q)
    \\
    &=
    \dfrac{2\alpha\alpha''}{\beta}
    +
    \dfrac{2\alpha^2\left(\alpha'\beta-\alpha\beta'\right)^2}{(\alpha^2+\tilde{\omega}^2)\beta^3}, 
  \end{split}
\end{align}
\begin{align}
  \begin{split}
    \dfrac{j_{\sin,1}^{(1)}(q)}{A_0}
    &=
    \dfrac{2\alpha^2\left(\alpha'\beta-\alpha\beta'\right)}{\beta^3}C_{1;\sin,1}(q)
    \\
    &=
    -\dfrac{2\alpha\left(\alpha'\beta-\alpha\beta'\right)^2\tilde{\omega}}{(\alpha^2+\tilde{\omega}^2)\beta^3}, 
  \end{split}
\end{align}
\begin{align}
  \begin{split}
    \dfrac{j_{\cos,2}^{(2)}(q)}{{A_0}^2}
    &=
    \dfrac{\alpha\alpha'''}{2\beta}
    +
    \dfrac{2\alpha^2\alpha''}{\beta^2}\dfrac{C_{1;\cos,1}(q)}{2}
    +
    \dfrac{\alpha^3\big(2\alpha'\beta-3\alpha\beta'\big)}{\beta^4}\dfrac{C_{1;\cos,1}(q)^2-C_{1;\sin,1}(q)^2}{2}
    +
    \dfrac{2\alpha^2\big(\alpha'\beta-\alpha\beta'\big)}{\beta^3}C_{2;\cos,2}(q), 
  \end{split}
\end{align}
\begin{align}
  \begin{split}
    \dfrac{j_{\sin,2}^{(2)}(q)}{{A_0}^2}
    &=
    \dfrac{2\alpha^2\alpha''}{\beta^2}\dfrac{C_{1;\sin,1}(q)}{2}
    +
    \dfrac{\alpha^3\big(2\alpha'\beta-3\alpha\beta'\big)}{\beta^4}C_{1;\cos,1}(q)C_{1;\sin,1}(q)
    +
    \dfrac{2\alpha^2\big(\alpha'\beta-\alpha\beta'\big)}{\beta^3}C_{2;\sin,2}(q), 
  \end{split}
\end{align}
\begin{align}
  \begin{split}
    \dfrac{j_{\cos,1}^{(3)}(q)}{{A_0}^3}
    &=
    \dfrac{\alpha^2\alpha'''}{\beta^2}\dfrac{3C_{1;\cos,1}(q)}{4}
    +
    \dfrac{2\alpha^3\alpha''}{\beta^3}
      \left(
        \dfrac{3C_{1;\cos,1}(q)^2}{4}+\dfrac{C_{1;\sin,1}(q)^2}{4}
      \right)
    +
    \dfrac{2\alpha^2\alpha''}{\beta^2}
      \left(
        C_{2;\mathrm{dc}}(q)+\dfrac{C_{2;\cos,2}(q)}{2}
      \right)
    \\
    &\quad
    +
    \dfrac{2\alpha^4\big(\alpha'\beta-2\alpha\beta'\big)}{\beta^5}
      \left(
        \dfrac{3C_{1;\cos,1}(q)^3}{4}+\dfrac{3C_{1;\cos,1}(q)C_{1;\sin,1}(q)^2}{4}
      \right)  
    \\
    &\quad
    +
    \dfrac{2\alpha^3\big(2\alpha'\beta-3\alpha\beta'\big)}{\beta^4}
      \left(
        C_{1;\cos,1}(q)C_{2;\mathrm{dc}}(q)
        +
        \dfrac{C_{1;\cos,1}(q)C_{2;\cos,2}(q)}{2}
        +
        \dfrac{C_{1;\sin,1}(q)C_{2;\sin,2}(q)}{2}
      \right)
    \\
    &\quad
    +
    \dfrac{2\alpha^2\big(\alpha'\beta-\alpha\beta'\big)}{\beta^3}C_{3;\cos,1}(q), 
  \end{split}
\end{align}
\begin{align}
  \begin{split}
    \dfrac{j_{\sin,1}^{(3)}(q)}{{A_0}^3}
    &=
    \dfrac{\alpha^2\alpha'''}{\beta^2}\dfrac{C_{1;\sin,1}(q)}{4}
    +
    \dfrac{2\alpha^3\alpha''}{\beta^3}\dfrac{C_{1;\cos,1}(q)C_{1;\sin,1}(q)}{2}
    +
    \dfrac{2\alpha^2\alpha''}{\beta^2}\dfrac{C_{2;\sin,2}(q)}{2}
    \\
    &\quad
    +
    \dfrac{2\alpha^4\big(\alpha'\beta-2\alpha\beta'\big)}{\beta^5}
      \left(
        \dfrac{3C_{1;\sin,1}(q)^3}{4}+\dfrac{3C_{1;\sin,1}(q)C_{1;\cos,1}(q)^2}{4}
      \right)  
    \\
    &\quad
    +
    \dfrac{2\alpha^3\big(2\alpha'\beta-3\alpha\beta'\big)}{\beta^4}
      \left(
        C_{1;\sin,1}(q)C_{2;\mathrm{dc}}(q)
        -
        \dfrac{C_{1;\sin,1}(q)C_{2;\cos,2}(q)}{2}
        +
        \dfrac{C_{1;\cos,1}(q)C_{2;\sin,2}(q)}{2}
      \right)
    \\
    &\quad
    +
    \dfrac{2\alpha^2\big(\alpha'\beta-\alpha\beta'\big)}{\beta^3}C_{3;\sin,1}(q), 
  \end{split}
\end{align}
\begin{align}
  \begin{split}
    \dfrac{j_{\cos,3}^{(3)}(q)}{{A_0}^3}
    &=
    \dfrac{\alpha^2\alpha'''}{\beta^2}\dfrac{C_{1;\cos,1}(q)}{4}
    +
    \dfrac{2\alpha^3\alpha''}{\beta^3}
      \left(
        \dfrac{C_{1;\cos,1}(q)^2}{4}-\dfrac{C_{1;\sin,1}(q)^2}{4}
      \right)
    +
    \dfrac{2\alpha^2\alpha''}{\beta^2}\dfrac{C_{2;\cos,2}(q)}{2}
    \\
    &\quad
    +
    \dfrac{2\alpha^4\big(\alpha'\beta-2\alpha\beta'\big)}{\beta^5}
      \left(
        \dfrac{C_{1;\cos,1}(q)^3}{4}-\dfrac{3C_{1;\cos,1}(q)C_{1;\sin,1}(q)^2}{4}
      \right)  
    \\
    &\quad
    +
    \dfrac{2\alpha^3\big(2\alpha'\beta-3\alpha\beta'\big)}{\beta^4}
      \left(
        \dfrac{C_{1;\cos,1}(q)C_{2;\cos,2}(q)}{2}
        -
        \dfrac{C_{1;\sin,1}(q)C_{2;\sin,2}(q)}{2}
      \right)
    \\
    &\quad
    +
    \dfrac{2\alpha^2\big(\alpha'\beta-\alpha\beta'\big)}{\beta^3}C_{3;\cos,3}(q), 
  \end{split}
\end{align}
\begin{align}
  \begin{split}
    \dfrac{j_{\sin,3}^{(3)}(q)}{{A_0}^3}
    &=
    \dfrac{\alpha^2\alpha'''}{\beta^2}\dfrac{C_{1;\sin,1}(q)}{4}
    +
    \dfrac{2\alpha^3\alpha''}{\beta^3}\dfrac{C_{1;\cos,1}(q)C_{1;\sin,1}(q)}{2}  
    +
    \dfrac{2\alpha^2\alpha''}{\beta^2}\dfrac{C_{2;\sin,2}(q)}{2}
    \\
    &\quad
    +
    \dfrac{2\alpha^4\big(\alpha'\beta-2\alpha\beta'\big)}{\beta^5}
      \left(
        -\dfrac{C_{1;\sin,1}(q)^3}{4}+\dfrac{3C_{1;\sin,1}(q)C_{1;\cos,1}(q)^2}{4}
      \right)  
    \\
    &\quad
    +
    \dfrac{2\alpha^3\big(2\alpha'\beta-3\alpha\beta'\big)}{\beta^4}
      \left(
        \dfrac{C_{1;\sin,1}(q)C_{2;\cos,2}(q)}{2}
        +
        \dfrac{C_{1;\cos,1}(q)C_{2;\sin,2}(q)}{2}
      \right)
    \\
    &\quad
    +
    \dfrac{2\alpha^2\big(\alpha'\beta-\alpha\beta'\big)}{\beta^3}C_{3;\sin,3}(q). 
  \end{split}
\end{align}
The validity of these results can be confirmed as follows. 
In the limit $\Gamma\rightarrow 0$, the TDGL equation leads to  
\begin{equation}
  \left.\psi(q,t)^2\right|_{\Gamma\rightarrow 0} = -\dfrac{\alpha(q+A(t))}{\beta(q+A(t))}, 
\end{equation}
and therefore 
\begin{equation}
  \left.f(q+A(t),\psi(q,t))\right|_{\Gamma\rightarrow 0}
  =-\dfrac{\alpha(q+A(t))^2}{2\beta(q+A(t))}
  =f(q+A(t)). 
\end{equation}
Using the relation 
\begin{equation}
  f(q+A(t))
  =f(q)
   +f'(q)A(t)
   +\dfrac{f''(q)}{2}A(t)^2
   +\dfrac{f'''(q)}{6}A(t)^3
   +O\left(A(t)^4\right), 
\end{equation}
and recalling $A(t)=A_0\cos\omega t$ with $A_0=\frac{2E_0}{\omega}$, where the factor of $2$ is included, we obtain 
\begin{align}
  \begin{split}
    \label{Eq:j(q+A(t))}
    j(q+A(t))
    &=-2\pdv{f(q+A(t))}{q}
    \\
    &=-2f'(q)
      -2f''(q)A(t)
      -f'''(q)A(t)^2
      -\dfrac{f''''(q)}{3}A(t)^3
      +O\left(A(t)^4\right)
    \\
    &=-2f'(q)
      -2f''(q)\cos\omega t\cdot A_0
      -\dfrac{f'''(q)}{2}\left(1+\cos 2\omega t\right)\cdot {A_0}^2
      -\dfrac{f''''(q)}{12}\left(3\cos\omega t+\cos 3\omega t\right)\cdot {A_0}^3
      +O\left({A_0}^4\right). 
  \end{split}
\end{align}
Here, we take the limit  $\left.j(q,t)\right|_{\Gamma\rightarrow 0}$ and compare it to Eq.~\eqref{Eq:j(q+A(t))}. 
Using 
\begin{equation}
  \left.C_{1;\cos,1}(q)\right|_{\Gamma\rightarrow 0}
  =\dfrac{\alpha'\beta-\alpha\beta'}{\alpha^2}, 
\end{equation}
\begin{equation}
  \left.C_{2;\mathrm{dc}}(q)\right|_{\Gamma\rightarrow 0}
  =\left.C_{2;\cos,2}(q)\right|_{\Gamma\rightarrow 0}
  =\dfrac{\alpha''\beta}{4\alpha^2}-\dfrac{\alpha'\big(\alpha'\beta-\alpha\beta'\big)}{2\alpha^3}, 
\end{equation}
\begin{equation}
  \dfrac{1}{3}\left.C_{3;\cos,1}(q)\right|_{\Gamma\rightarrow 0}
  =\left.C_{3;\cos,3}(q)\right|_{\Gamma\rightarrow 0}
  =\dfrac{\alpha'''\beta}{24\alpha^2}
   -\dfrac{\alpha'\alpha''\beta}{8\alpha^3}
   -\dfrac{\alpha''\big(\alpha'\beta-\alpha\beta'\big)}{8\alpha^3}
   +\dfrac{\alpha'^2\big(\alpha'\beta-\alpha\beta'\big)}{4\alpha^4}, 
\end{equation}
and 
\begin{equation}
  \left.C_{1;\sin,1}(q)\right|_{\Gamma\rightarrow 0}=0,\quad
  \left.C_{2;\sin,2}(q)\right|_{\Gamma\rightarrow 0}=0,\quad
  \left.C_{3;\sin,1}(q)\right|_{\Gamma\rightarrow 0}=0,\quad
  \left.C_{3;\sin,3}(q)\right|_{\Gamma\rightarrow 0}=0, 
\end{equation}
we obtain 
\begin{equation}
  \left.\dfrac{j_{\cos,1}^{(1)}(q)}{A_0}\right|_{\Gamma\rightarrow 0}
  =\dfrac{2\alpha\alpha''}{\beta}
   +\dfrac{2\big(\alpha'\beta-\alpha\beta'\big)^2}{\beta^3}
  =-2f''(q), 
\end{equation}
\begin{equation}
  \left.\dfrac{j_{\mathrm{photo}}^{(2)}(q)}{{A_0}^2}\right|_{\Gamma\rightarrow 0}
  =\left.\dfrac{j_{\cos,2}^{(2)}(q)}{{A_0}^2}\right|_{\Gamma\rightarrow 0}
  =\dfrac{\alpha\alpha'''}{2\beta}
   +\dfrac{3\alpha''\big(\alpha'\beta-\alpha\beta'\big)}{2\beta^2}
   -\dfrac{3\beta'\big(\alpha'\beta-\alpha\beta'\big)^2}{2\beta^4}
  =-\dfrac{f'''(q)}{2}, 
\end{equation}
\begin{equation}
  \dfrac{1}{3}\left.\dfrac{j_{\cos,1}^{(3)}(q)}{{A_0}^3}\right|_{\Gamma\rightarrow 0}
  =\left.\dfrac{j_{\cos,3}^{(3)}(q)}{{A_0}^3}\right|_{\Gamma\rightarrow 0}
  =\dfrac{\alpha''^2}{4\beta}
   +\dfrac{\alpha'''\big(\alpha'\beta-\alpha\beta'\big)}{3\beta^2}
   -\dfrac{\alpha''\beta'\big(\alpha'\beta-\alpha\beta'\big)}{\beta^3}
   +\dfrac{\beta'^2\big(\alpha'\beta-\alpha\beta'\big)^2}{\beta^5}
  =-\dfrac{f''''(q)}{12}, 
\end{equation}
and 
\begin{equation}
  \left.\dfrac{j_{\sin,1}^{(1)}(q)}{A_0}\right|_{\Gamma\rightarrow 0}=0,\quad
  \left.\dfrac{j_{\sin,2}^{(2)}(q)}{{A_0}^2}\right|_{\Gamma\rightarrow 0}=0,\quad
  \left.\dfrac{j_{\sin,1}^{(3)}(q)}{{A_0}^3}\right|_{\Gamma\rightarrow 0}=0,\quad
  \left.\dfrac{j_{\sin,3}^{(3)}(q)}{{A_0}^3}\right|_{\Gamma\rightarrow 0}=0. 
\end{equation}
These results confirm the validity of the approximate solutions in the limit $\Gamma\rightarrow 0$.

\section{Comparison between the numerical solution and the approximate solution}

\begin{figure}[H]
  \centering
  \includegraphics[width=0.7\columnwidth]{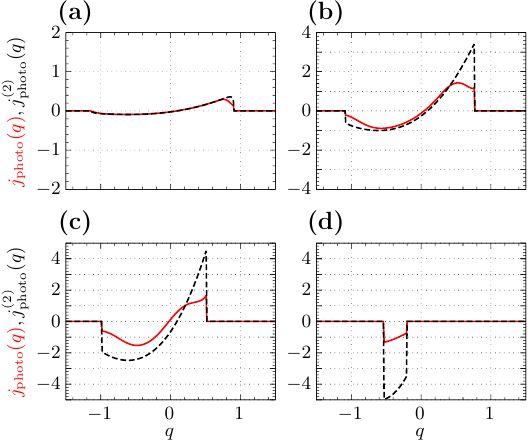}
  \caption{
    Comparison of the numerical solution $j_{\mathrm{photo}}(q)$ (red line) with the approximate solution $j_{\mathrm{photo}}^{(2)}(q)\propto{E_0}^2$ (black dashed line). 
    Panels (a), (b), (c), and (d) show the results obtained for $E_0=0.01$, $0.033$, $0.052$, and $0.074$, respectively. 
    Panel (d) lies in the perfect SDE regime. 
  }
  \label{FigS3}
\end{figure}

Focusing on the case of monochromatic light, we compare the numerical solution of the photocurrent $j_{\mathrm{photo}}(q)$ with the second-order photocurrent $j_{\mathrm{photo}}^{(2)}(q)\propto{E_0}^2$ obtained from the approximate solution. 
We set $\alpha_3=0.2$, $\beta_1=0$, and $\omega=0.1$. 
Figure~\ref{FigS3} shows the results for various amplitudes of the ac electric field $E_0$. 
Panel (a) corresponds to the small $E_0$ regime ($E_0=0.01$), (b) and (c) to the intermediate $E_0$ regime ($E_0=0.033$ and $0.052$), and (d) to the large $E_0$ regime ($E_0=0.074$) where the perfect SDE emerges. 
As $E_0$ increases, the difference between the numerical solution and the approximate solution becomes more pronounced. 
Importantly, the photo-induced dc supercurrent in the intermediate and large $E_0$ regimes, including the region where the perfect SDE is realized, cannot be explained solely by the second-order optical response $\left(\propto{E_0}^2\right)$. 
This indicates that higher even-order responses play an essential role in the realization of the perfect SDE. 

\section{Coupling between Higgs modes and electromagnetic fields}

We have focused on the approximate solution for the supercurrent $j(q,t)$ so far. 
Here, we discuss the approximate solution for the order parameter $\psi(q,t)$, which describes the coupling between electromagnetic fields and the amplitude mode, known as the Higgs mode. 
From Eq.~\eqref{Eq:psi^2}, we obtain 
\begin{align}
  \begin{split}
    \psi(q,t)
    &\sim
    \sqrt{\dfrac{1}{C_0(q)}}
    \left[
      1
      +
      \dfrac{\tilde{C}_1(q,t)}{2}A_0
      +
      \left(
        -\dfrac{\tilde{C}_1(q,t)^2}{8}
        +\dfrac{\tilde{C}_2(q,t)}{2}
      \right){A_0}^2
      +
      \left(
        \dfrac{\tilde{C}_1(q,t)^3}{16}
        -\dfrac{\tilde{C}_1(q,t)\tilde{C}_2(q,t)}{4}
        +\dfrac{\tilde{C}_3(q,t)}{2}
      \right){A_0}^3
    \right]
    \\
    &=
    \sqrt{\dfrac{1}{C_0(q)}}
    \Bigg[
      1
      +
      \dfrac{\alpha(q)}{2\beta(q)}C_1(q,t)A_0
      +
      \left(
        \dfrac{3\alpha(q)^2}{8\beta(q)^2}C_1(q,t)^2
        +\dfrac{\alpha(q)}{2\beta(q)}C_2(q,t)
      \right){A_0}^2
      \\
      &\qquad\qquad\qquad\qquad
      +
      \left(
        \dfrac{5\alpha(q)^3}{16\beta(q)^3}C_1(q,t)^3
        +\dfrac{3\alpha(q)^2}{4\beta(q)^2}C_1(q,t)C_2(q,t)
        +\dfrac{\alpha(q)}{2\beta(q)}C_3(q,t)
      \right){A_0}^3
    \Bigg]
    \\
    &=\psi^{(0)}(q)\big(1+\psi^{(1)}(q,t)+\psi^{(2)}(q,t)+\psi^{(3)}(q,t)\big). 
  \end{split}
\end{align}
Here, $\psi^{(n)}(q,t)\propto {A_0}^n$, and 
we obtain 
\begin{equation}
  \psi^{(0)}(q)=\sqrt{-\dfrac{\alpha(q)}{\beta(q)}}, 
\end{equation}
\begin{equation}
  \psi^{(1)}(q,t)
  =\psi_{\cos,1}^{(1)}(q)\cos\omega t + \psi_{\sin,1}^{(1)}(q)\sin\omega t, 
\end{equation}
\begin{equation}
  \psi^{(2)}(q,t)
  =\psi_{\mathrm{photo}}^{(2)}(q) + \psi_{\cos,2}^{(2)}(q)\cos 2\omega t + \psi_{\sin,2}^{(2)}(q)\sin 2\omega t, 
\end{equation}
\begin{equation}
  \psi^{(3)}(q,t)
  =\psi_{\cos,1}^{(3)}(q)\cos\omega t + \psi_{\sin,1}^{(3)}(q)\sin\omega t + \psi_{\cos,3}^{(3)}(q)\cos 3\omega t + \psi_{\sin,3}^{(3)}(q)\sin 3\omega t, 
\end{equation}
where
\begin{equation}
  \dfrac{\psi_{\cos,1}^{(1)}(q)}{A_0}
  =\dfrac{\alpha(q)}{2\beta(q)}C_{1;\cos,1}(q), 
\end{equation}
\begin{equation}
  \dfrac{\psi_{\sin,1}^{(1)}(q)}{A_0}
  =\dfrac{\alpha(q)}{2\beta(q)}C_{1;\sin,1}(q), 
\end{equation}
\begin{equation}
  \dfrac{\psi_{\mathrm{photo}}^{(2)}(q)}{{A_0}^2}
  =\dfrac{3\alpha(q)^2}{8\beta(q)^2}\dfrac{C_{1;\cos,1}(q)^2+C_{1;\sin,1}(q)^2}{2}
  +\dfrac{\alpha(q)}{2\beta(q)}C_{2;\mathrm{dc}}(q), 
\end{equation}
\begin{equation}
  \dfrac{\psi_{\cos,2}^{(2)}(q)}{{A_0}^2}
  =\dfrac{3\alpha(q)^2}{8\beta(q)^2}\dfrac{C_{1;\cos,1}(q)^2-C_{1;\sin,1}(q)^2}{2}
  +\dfrac{\alpha(q)}{2\beta(q)}C_{2;\cos,2}(q), 
\end{equation}
\begin{equation}
  \dfrac{\psi_{\sin,2}^{(2)}(q)}{{A_0}^2}
  =\dfrac{3\alpha(q)^2}{8\beta(q)^2}C_{1;\cos,1}(q)C_{1;\sin,1}(q)
  +\dfrac{\alpha(q)}{2\beta(q)}C_{2;\sin,2}(q), 
\end{equation}
\begin{align}
  \begin{split}
    \dfrac{\psi_{\cos,1}^{(3)}(q)}{{A_0}^3}
    &=\dfrac{5\alpha(q)^3}{16\beta(q)^3}
      \left(
        \dfrac{3C_{1;\cos,1}(q)^3}{4}+\dfrac{3C_{1;\cos,1}(q)C_{1;\sin,1}(q)^2}{4}
      \right)
    \\
    &\quad
    +\dfrac{3\alpha(q)^2}{4\beta(q)^2}
      \left(
        C_{1;\cos,1}(q)C_{2;\mathrm{dc}}(q)
        +
        \dfrac{C_{1;\cos,1}(q)C_{2;\cos,2}(q)}{2}
        +
        \dfrac{C_{1;\sin,1}(q)C_{2;\sin,2}(q)}{2}
      \right)
    +\dfrac{\alpha(q)}{2\beta(q)}C_{3;\cos,1}(q), 
  \end{split}
\end{align}
\begin{align}
  \begin{split}
    \dfrac{\psi_{\sin,1}^{(3)}(q)}{{A_0}^3}
    &=\dfrac{5\alpha(q)^3}{16\beta(q)^3}
      \left(
        \dfrac{3C_{1;\sin,1}(q)^3}{4}+\dfrac{3C_{1;\sin,1}(q)C_{1;\cos,1}(q)^2}{4}
      \right)
    \\
    &\quad
    +\dfrac{3\alpha(q)^2}{4\beta(q)^2}
      \left(
        C_{1;\sin,1}(q)C_{2;\mathrm{dc}}(q)
        -
        \dfrac{C_{1;\sin,1}(q)C_{2;\cos,2}(q)}{2}
        +
        \dfrac{C_{1;\cos,1}(q)C_{2;\sin,2}(q)}{2}
      \right)
    +\dfrac{\alpha(q)}{2\beta(q)}C_{3;\sin,1}(q), 
  \end{split}
\end{align}
\begin{align}
  \begin{split}
    \dfrac{\psi_{\cos,3}^{(3)}(q)}{{A_0}^3}
    &=\dfrac{5\alpha(q)^3}{16\beta(q)^3}
      \left(
        \dfrac{C_{1;\cos,1}(q)^3}{4}-\dfrac{3C_{1;\cos,1}(q)C_{1;\sin,1}(q)^2}{4}
      \right)
    \\
    &\quad
    +\dfrac{3\alpha(q)^2}{4\beta(q)^2}
      \left(
        \dfrac{C_{1;\cos,1}(q)C_{2;\cos,2}(q)}{2}
        -
        \dfrac{C_{1;\sin,1}(q)C_{2;\sin,2}(q)}{2}
      \right)
    +\dfrac{\alpha(q)}{2\beta(q)}C_{3;\cos,3}(q), 
  \end{split}
\end{align}
\begin{align}
  \begin{split}
    \dfrac{\psi_{\sin,3}^{(3)}(q)}{{A_0}^3}
    &=\dfrac{5\alpha(q)^3}{16\beta(q)^3}
      \left(
        -\dfrac{C_{1;\sin,1}(q)^3}{4}+\dfrac{3C_{1;\sin,1}(q)C_{1;\cos,1}(q)^2}{4}
      \right)
    \\
    &\quad
    +\dfrac{3\alpha(q)^2}{4\beta(q)^2}
      \left(
        \dfrac{C_{1;\sin,1}(q)C_{2;\cos,2}(q)}{2}
        +
        \dfrac{C_{1;\cos,1}(q)C_{2;\sin,2}(q)}{2}
      \right)
    +\dfrac{\alpha(q)}{2\beta(q)}C_{3;\sin,3}(q). 
  \end{split}
\end{align}
For the GL models where only $\alpha_0$, $\alpha_2$, and $\beta_0$ are nonzero, these results are consistent with the understanding of conventional superconductivity. 
Indeed, we obtain $\psi^{(1)}(0,t)\equiv 0$, indicating that the Higgs mode does not respond linearly to the electromagnetic field~\cite{Shimano2020}. 
In contrast, when symmetry is broken by current injection, $\psi^{(1)}(q,t)\not\equiv 0$ for $q\neq 0$ indicates that the linear response is allowed~\cite{Moor2017, Nakamura2019}.

\section{Approximate solution under two-frequency light irradiation}

We consider two-frequency light with frequencies $\omega$ and $2\omega$, described by the $x$-component of the vector potential $A(t)=A_1\cos\omega t+A_2\cos 2\omega t$, where $A_n=\frac{2E_n}{n\omega}$. 
As we have done in deriving the approximate solution for monochromatic light, we incorporate the factor of $2$ in the argument of the GL coefficients into $A_n$ for notational simplicity. 
Following the same procedure as in the monochromatic case, we obtain an approximate solution of the supercurrent. 
For systems with inversion or time-reversal symmetry, we take $\alpha(q)=\alpha_0+\alpha_2q^2$ and $\beta(q)=\beta_0$. 
The second-order photocurrent $j_{\mathrm{photo}}^{(2)}(q)$ is obtained as: 
\begin{equation}
  j_{\mathrm{photo}}^{(2)}(q)
  =
    \dfrac{\alpha'\alpha''(3\alpha^2+\tilde{\omega}^2)}{2(\alpha^2+\tilde{\omega}^2)\beta}
  {A_1}^2
  +
    \dfrac{\alpha'\alpha''(3\alpha^2+4\tilde{\omega}^2)}{2(\alpha^2+4\tilde{\omega}^2)\beta}
  {A_2}^2, 
\end{equation}
where the argument $q$ of the GL coefficients $\alpha(q)$ and $\beta(q)$ is omitted. 
This formula is a straightforward extension of the formula for monochromatic light. 
A crucial difference from the case of monochromatic light is the presence of a dc supercurrent arising from the third-order optical response. 
The third-order photocurrent $j_{\mathrm{photo}}^{(3)}(q)$ is given by 
\begin{equation}
  j_{\mathrm{photo}}^{(3)}(q)
  =
  \left[
    \dfrac{3\alpha^2\alpha''^2(\alpha^2+3\tilde{\omega}^2)}
    {4(\alpha^2+\tilde{\omega}^2)(\alpha^2+4\tilde{\omega}^2)\beta}
    +
    \dfrac{3\alpha^3\alpha'^2\alpha''\tilde{\omega}^2}
    {(\alpha^2+\tilde{\omega}^2)^2(\alpha^2+4\tilde{\omega}^2)\beta}
  \right]
  {A_1}^2A_2. 
\end{equation}
In the limit $\omega\rightarrow 0$, we obtain the asymptotic form 
\begin{equation}
\label{Eq:j_{photo}^{(3)}}
  j_{\mathrm{photo}}^{(3)}(q)
  \rightarrow
  -\dfrac{\rho_{\mathrm{NLSF}}(q)}{6}
  \cdot\dfrac{3}{4}
  \cdot\left(
    \dfrac{E_1}{\omega}
  \right)^2
  \cdot\dfrac{E_2}{2\omega}
  \cdot\theta(-\gamma_0(q)), 
\end{equation}
where the nonlinear superfluid density (NLSF) is defined by $\rho_{\mathrm{NLSF}}(q)\coloneq{\partial_A}^4f_{\mathrm{eq}}(q+2A)|_{A=0}$, as in the main text. 